\documentclass[]{spie}  %>>> use for US letter paper
\usepackage{xcolor}
\usepackage{graphicx}
\usepackage[space]{grffile} % extends the file name processing of package graphics

\usepackage{amsfonts,amsmath,amssymb}

\usepackage[utf8]{inputenc}
\usepackage[colorlinks=true, allcolors=blue]{hyperref}

\usepackage{textcomp}
\usepackage{longtable}
\usepackage{multirow,booktabs}

\newcommand*{\ROW}{} % predefined here, redefined later
\newcommand*{\ROWTITLE}{} % predefined here, redefined later

\newcommand{\KEYWORD}[1]{\texttt{#1}} % FITS keyword
\newcommand{\STRING}[1]{\texttt{'#1'}} % FITS string value
\newcommand{\VARIABLE}[1]{\texttt{\textsl{#1}}}
\title{IMAGE-OI: an OIFITS extension and its application in OImaging to compare image reconstruction algorithms}

\author[a]{Ferréol Soulez}
\author[b]{Laurent Bourgès}
\author[a]{Antoine Kaszczyc}
\author[b]{Guillaume Mella}
\author[a]{Martin Pratoussy}
\author[b]{Gilles Duvert}
\author[c]{Jacques Kluska}
\author[a]{Eric Thiébaut}
\author[d]{John Young}
\affil[a]{Univ. Lyon, Univ. Lyon 1, ENS de Lyon, CNRS, Centre de Recherche Astrophysique de Lyon UMR5574, F-69230, Saint-Genis-Laval, France}
\affil[b]{Univ. Grenoble Alpes, CNRS, Observatoire des Sciences de l'Univers de Grenoble (France)}
\affil[c]{Institute of Astronomy, KU Leuven (Belgium)}
\affil[d]{University of Cambridge, Cambridge (UK)}

\authorinfo{Further author information: E-mail: ferreol.soulez@univ-lyon1.fr}

\begin{document}
\maketitle

\begin{abstract}
	In interferometry, the quality of the reconstructed image depends on the algorithm used and its parameters, and users often need to compare the results of several algorithms to disentangle artifacts from actual features of the astrophysical object. Such comparisons can rapidly become cumbersome, as these software packages are very different. OImaging is a graphical interface intended to be a common frontend to image reconstruction software packages. With OImaging, the user can now perform multiple reconstructions within a single interface. From a given dataset, OImaging allows benchmarking of different image reconstruction algorithms and assessment of the reliability of the image reconstruction process. To that end, OImaging uses the IMAGE-OI OIFITS extension proposed to standardize communication with image reconstruction algorithms.
\end{abstract}


\section{INTRODUCTION}
\label{sec:intro}  % \label{} allows reference to this section

Many image reconstruction algorithms (see \cite{thiebaut2017principles} for a review) have been developed to process optical
interferometric data.  Using these algorithms may require substantial
expertise (for instance you may have to learn a specific programming
language). To make image reconstruction accessible to as many people as possible, we have developed a common
graphical user interface (GUI) to drive such imaging algorithms in a
\emph{user friendly} way. For obvious reasons, we also wanted to avoid, as far as possible, rewriting
the image reconstruction algorithms.  

The purpose of this work is twofold: (i) proposes an \textbf{IMAGE-OI OI-FITS extension} to communicate with image reconstruction software backends and (ii) presents \textbf{OImaging}, a graphical user interface that uses this standard to provide a common frontend for several image reconstruction algorithms.

\section{Image-OI: a common interface for image reconstruction software packages}
\label{sec:OI-image}

%\subsection{Architecture?}

All existing algorithms take their input data in the OIFITS (either v1 or v2) format
\cite{Pauls_et_al-2005-oifits,Duvert_et_al-2017-OIFITS2}.  This format stores the optical
interferometric data in FITS binary tables.  More generally, a FITS file
\cite{Pence_et_al-2010-FITS} comprises a number of so-called header data
units (HDU), each HDU having a header part with scalar parameters
specified by FITS keywords\footnote{A FITS keyword consists of upper case
  latin letters, digits, hyphen or underscore characters and has at least
  one character and at most 8 characters.} and a data part.  The
header part obeys a specific format but it is textual and easy to read for
an human.  The data part usually consists of binary data stored in various
forms.  We only consider FITS \emph{images} (that is a
multidimensional array of values of the same data type) and FITS
\emph{binary tables}.  These are sufficient for our purposes.

We are currently using an extension of the OIFITS format to build a common software interface
for image reconstruction algorithms \footnote{further discussions about IMAGE-OI can be found at \url{https://github.com/JMMC-OpenDev/OI-Imaging-JRA}}. This extension will soon be proposed to the OIFITS group for approval.   
By adding FITS HDUs or introducing new FITS keywords, it is possible to
provide any additional data (\emph{e.g.} the initial image) and
parameters required to run the image reconstruction process. 
Thus, the input file is a valid OIFITS file (resulting
from the merging of all the optical interferometric data to process)
with  additional HDUs containing further data needed for the
reconstruction (for instance, the initial image) and with scalar input
settings provided as the values of FITS keywords.  This is detailed in
the following subsections.

%Note that the FITS standard states that angular measurements expressed as floating-point values and specified with reserved keywords \emph{should} be given in degrees.

The HDUs that are specific to the interface specification described in this
paper have \KEYWORD{EXTNAME} values prefixed with \STRING{IMAGE-OI}.
This is to distinguish them from OIFITS HDUs, which use the prefix
\STRING{OI\_}. At least three of these HDUs must be added:
\begin{itemize}
	\item the image \STRING{IMAGE-OI~INITIAL~IMAGE} that contains the initial image guess required to start the algorithm,
	\item the binary table \STRING{IMAGE-OI~INPUT~PARAM} that contains all input parameters needed for the reconstruction, and
	\item the binary table \STRING{IMAGE-OI~OUTPUT~PARAM} that contains all output parameters given by the reconstruction algorithm.
\end{itemize}
A command line tool for creating and editing these IMAGE-OI OIFITS files has been developed,
see \url{https://github.com/jsy1001/OI-Interface-Tools}.

\subsection{Input parameters}

All input parameters needed by the reconstruction algorithms must be stored in a separate HDU.
In anticipation of perhaps needing to store vector parameters in future,
this HDU shall be a binary table HDU.  This binary table shall have the
\KEYWORD{EXTNAME} keyword set to \STRING{IMAGE-OI~INPUT~PARAM} and shall
contain all of the non-image parameters, with the exception of the pixel
size and the dimensions of the reconstructed image which are specified by
those of the initial image (see below), itself stored in a dedicated
(image) HDU. Table~\ref{tab:input-params} gives the minimal list of  input
parameters required by reconstruction algorithms. Reconstruction algorithms shouldn't modify this HDU. If 
the actual value of an input parameter used by a algorithm changes, the new value must be given in \STRING{IMAGE-OI OUTPUT PARAM}.

Note that the definition of this \STRING{IMAGE-OI OUTPUT PARAM} HDU is sufficiently versatile to also store parameters used by model fitting algorithms.

%==============================================================================
\renewcommand{\ROW}[3]{\KEYWORD{#1} & #2 &#3 \\}
\renewcommand{\ROWTITLE}[1]{\multicolumn{3}{c}{\textbf{#1}} \\*
  Keyword & Type & Description \\*}

\begin{longtable}[c]{lcp{83mm}}
  %\centering
  \caption[FITS keywords]{FITS keywords used to specify the input
  parameters in header of the binary table HDU
  with \KEYWORD{EXTNAME} of \STRING{IMAGE-OI~INPUT~PARAM}.
  \label{tab:input-params}}
  \endfirsthead
  \caption[]{(continued)}
  \endhead
  \hline
  \ROWTITLE{Data Selection (mandatory)}
  \hline
  \ROW{TARGET}{string}{Identifier of the target object to reconstruct}
  \ROW{WAVE\_MIN}{real}{Minimum wavelength to select (in meters)}
  \ROW{WAVE\_MAX}{real}{Maximum wavelength to select (in meters)}
  \ROW{USE\_VIS}{string}{Complex visibility data to consider if any$^\dagger$}*
  \ROW{USE\_VIS2}{logical}{Use squared visibility data if any}
  \ROW{USE\_T3}{string}{Bispectrum data to consider if any$^\dagger$}[1ex]
  \multicolumn{3}{c}{$^\dagger$ value can be: \STRING{NONE}, \STRING{ALL},
   \STRING{AMP} or \STRING{PHI}.}\\
  \hline
  \ROWTITLE{Algorithm Settings}
  \hline
  \ROW{INIT\_IMG}{string}{Identifier of the initial image}
  \ROW{MAXITER}{integer}{Maximum number of iterations to run}
  \ROW{RGL\_NAME}{string}{Name of the regularization method}
  \ROW{AUTO\_WGT}{logical}{Automatic regularization weight}
  \ROW{RGL\_WGT}{real}{Weight of the regularization}
  \ROW{RGL\_PRIO}{string}{Identifier of the HDU with the prior image}
  \ROW{FLUX}{real}{Assumed total flux (1 is the default)}
  \ROW{FLUXERR}{real}{Error bar for the total flux (0 means strict constraint)}
  \ROW{HDUPREFX}{string}{Prefix that specifies the leading text to use in the HDUNAME of the final image}
  \hline
\end{longtable}

\subsubsection{Data selection}

To keep things simple, any sophisticated selection, merging or editing of
the data should be done by a separate tool.  The image reconstruction
software applications shall assume that they receive clean input data.
There are however a few parameters devoted to the selection of data.  The
\KEYWORD{TARGET} keyword specifies the name of the target object to
reconstruct.  The value of this keyword should match one of the identifiers
in the column \KEYWORD{TARGET} in the \texttt{OI\_TARGET} binary table of
the OIFITS file.  In order to restrict the types of interferometric data
used for the reconstruction, keywords \KEYWORD{USE\_VIS},
\KEYWORD{USE\_VIS2} and \KEYWORD{USE\_T3} should be set with values
specifying which complex visibility data, powerspectrum data and bispectrum
data to use if any.  More specifically, keywords \KEYWORD{USE\_VIS} and
\KEYWORD{USE\_T3} take string values which are \STRING{NONE}, \STRING{ALL},
\STRING{AMP} or \STRING{PHI} to indicate whether all, none, only the
amplitude or only the phase of such data are to be used.  Keyword
\KEYWORD{USE\_VIS2} has a boolean value indicating whether to use the
powerspectrum data.  Not all algorithms can use all types of data and the
values of these keywords should be set (in the output file) to
reflect what was really used.

The primary objective is to consider monochromatic image reconstruction.
The interferometric data are generally available at many wavelengths.  The
result will, in fact, be a gray image of the target built from the data in
the wavelength range (inclusive) specified by the FITS keywords
\KEYWORD{WAVE\_MIN} and \KEYWORD{WAVE\_MAX}. The wavelength range is given
in meters.

Whatever these settings, the image reconstruction algorithms must
honor the \KEYWORD{FLAG} column of the OI data.  We recall that this
field has a logical value which is true when the corresponding piece
of data should be discarded (in addition to any data with NULL values) and false when it should be considered.

\subsubsection{Algorithm settings}

Any parameter needed by the reconstruction algorithm can be added as a keyword in the  \STRING{IMAGE-OI~INPUT~PARAM} HDU paying attention to the naming constraints of FITS keywords. The algorithm settings keywords shown in table~\ref{tab:input-params} are common to several algorithms (if not all as e.g. \KEYWORD{MAXITER}). More generally, even for future keywords, algorithms should use the same keywords for common parameters.  The types of these keywords can be logical, integer, real or string. A string keyword can be used to identify an HDU containing additional information. For example, \KEYWORD{RGL\_PRIO} contains the identifier of the HDU with the prior image needed by the reconstruction algorithm.

\subsection{Initial image}
\label{sec:initial}

All reconstruction algorithms are iterative and require an initial
image to start with. The initial image is provided as a FITS image in
one of the HDUs of the input file. The primary HDU can be used to store
the initial image but in that case it has to be moved to another location in the output OIFITS file. The pixel size and the dimensions of the
initial image determine those of the reconstructed image.  The
dimensions are given by the FITS keywords \KEYWORD{NAXIS1} and
\KEYWORD{NAXIS2} while the pixel size is given by the FITS keywords
\KEYWORD{CDELT1} and \KEYWORD{CDELT2} (both values must have the same magnitude but may differ in sign).

It is not intended that the image reconstruction algorithm be able to deal
with any possible world coordinate system (WCS) nor with any possible
coordinate units.  We therefore restrict this standard to WCS which can be
fully specified by the \KEYWORD{CRPIX\VARIABLE{i}},
\KEYWORD{CRVAL\VARIABLE{i}}, \KEYWORD{CDELT\VARIABLE{i}},
\KEYWORD{CTYPE\VARIABLE{i}}, and \KEYWORD{CUNIT\VARIABLE{i}} keywords
(where \VARIABLE{i} is the axis number). Other WCS keywords like
\KEYWORD{CROTAn}, \KEYWORD{PC\VARIABLE{i}\_\VARIABLE{j}},
\KEYWORD{DC\VARIABLE{i}\_\VARIABLE{j}}, should not be specified as their
default values (according to the FITS standard) are not suitable for us.  The
same WCS conventions hold for any output image produced by the
reconstruction algorithm.  For any external software packages to correctly display
the images, the parameters of the WCS must be completely and correctly
specified. On the sub-arcsecond fields of view of optical interferometry observations all projections are for all purposes identical.

The conventions are that the first image axis corresponds to right
ascension (RA) and second image axis corresponds to declination (DEC) both
relative to the center of the field of view (FOV) specified by the keywords
\KEYWORD{CRPIX\VARIABLE{i}} (in fractional pixel units). If keywords
\KEYWORD{CRPIX\VARIABLE{i}} are omitted, they default to the geometric
center of the FOV. The pixel size (which is specified by the absolute
values of \KEYWORD{CDELT1} and \KEYWORD{CDELT2}) must be the same in both
directions.  Following standard conventions for display of a celestial
image, to have the relative right ascension (RA) oriented toward East to
correspond to the left (first columns) of the image, \KEYWORD{CDELT1}
should be strictly negative, while, to have the relative declination (DEC)
oriented toward North to correspond to the top (last rows) of the image,
\KEYWORD{CDELT2} should be strictly positive. By default, the physical
coordinate units are in degrees; otherwise \KEYWORD{CUNIT\VARIABLE{i}} may
be \verb+'deg'+ for degrees or \verb+'arcsec'+ for arcseconds.
Table~\ref{tab:initial-image} summarizes these rules.

The OIFITS norm specifies that \KEYWORD{RAEP0} and \KEYWORD{DECEP0}
in \KEYWORD{OI\_TARGET} are the coordinates of the phase center, which
we assume to be the absolute world coordinates of the
reference pixel in the reconstructed image (Notes that if the data is not phase referenced, the reconstructed image is shift invariant and the absolute coordinates are unknown).

%==============================================================================
\renewcommand{\ROW}[3]{\KEYWORD{#1} & #2 &#3 \\}
\renewcommand{\ROWTITLE}[1]{\multicolumn{3}{c}{\textbf{#1}} \\*
  Keyword & Type & Description \\*}

\begin{longtable}[c]{lcp{83mm}}
  %\centering
  \caption[FITS keywords]{Initial image HDU keywords. The \KEYWORD{HDUNAME} of this HDU is
  specified by the value of the \KEYWORD{INIT\_IMG} keyword in  \STRING{IMAGE-OI~INPUT~PARAM}.
  \label{tab:initial-image}}
  \endhead
  \ROWTITLE{Image Parameters}
  \hline
  \ROW{HDUNAME}{string}{Unique name for the image within the FITS file}
  \ROW{NAXIS1}{integer}{First dimension of the image}
  \ROW{NAXIS2}{integer}{Second dimension of the image}
  \ROW{CTYPE1}{string}{Coordinate name \STRING{RA---TAN} for 1st axis}
  \ROW{CTYPE2}{string}{Coordinate name \STRING{DEC--TAN} for 2nd axis}
  \ROW{CDELT{\slshape i}}{real}{Physical increment along \texttt{\slshape i}-th
    dimension of the image (for \texttt{\slshape i} = 1 or 2)}
  \ROW{CUNIT{\slshape i}}{string}{Physical units for \KEYWORD{CDELT{\slshape i}}
    and \KEYWORD{CRVAL{\slshape i}}; defaults to \STRING{deg} if omitted}
  \ROW{CRPIX{\slshape i}}{real}{Index of reference pixel along \texttt{\slshape i}-th
    dimension (for \texttt{\slshape i} = 1 or 2); defaults to the geometric center
    of the field of view if omitted}
  \ROW{CRVAL{\slshape i}}{real}{Physical coordinate of reference pixel along
    \texttt{\slshape i}-th dimension (for \texttt{\slshape i} = 1 or 2) and
    relative to the center of the field of view; defaults to 0 if omitted}
  \hline
\end{longtable}

\subsection{Output parameters}

It is proposed that the output format be as similar as possible as the input
format.  The output file must provide the reconstructed image but also some
information for interpreting the result.  

Any scalar output parameters from the image reconstruction shall be
stored in a binary table HDU with \KEYWORD{EXTNAME} of
\STRING{IMAGE-OI OUTPUT PARAM}. Examples of scalar outputs include
$\chi^2$, any regularization parameters estimated from the data, and
any model parameters that are not image pixels (for example stellar
disk parameters in SPARCO\cite{kluska2014sparco}). If an algorithm automatically changes some input parameters 
(e.g. \KEYWORD{UVMAX} with \cite{baron2008BSMEM}), the new value should be written in this \STRING{IMAGE-OI OUTPUT PARAM} leaving \STRING{IMAGE-OI INPUT PARAM} untouched.

Standard output parameters are listed in
Table~\ref{tab:output-params}.
\newpage

\begin{longtable}[c]{lcp{90mm}}
  %\centering
  \caption[FITS keywords]{FITS keywords used to specify the output
  parameters. These keywords must be stored in a binary table HDU
  with \KEYWORD{EXTNAME} of \STRING{IMAGE-OI~OUTPUT~PARAM}.
  \label{tab:output-params}}
  \endfirsthead
  \caption[]{(continued)}
  \endhead
  \hline
  \ROWTITLE{Algorithm Results}
  \hline
  \ROW{LAST\_IMG}{string}{Identifier of the final image}
  \ROW{NITER}{integer}{Total iterations done in the current program run}
  \ROW{CHISQ}{real}{Reduced chi-squared}
  \ROW{FPRIOR}{real}{Regularization penalty}
  \ROW{FLUX}{real}{Total image flux}
  \ROW{PROCSOFT}{string}{Software name and version number}
  \ROW{CONVERGE}{boolean}{Set to 'T' if the software stopped because it has converged}
  \hline
\end{longtable}

\subsection{Final image}

To compare the initial and the final images, they must be stored in
different HDUs, hence the FITS keyword \KEYWORD{HDUNAME} is used to
distinguish them (the EXTNAME keyword should not be used in the
primary HDU). As explained previously, the dimensions, pixel size and
orientation of the output image(s) are the same as for the initial image.

As most image viewers
are only capable of displaying the image stored in the primary HDU, we
suggest storing the initial image in the primary HDU of the input
file, but storing the final or current image in the primary HDU of the
output file.  The idea is to have the most relevant image stored in
the primary HDU.  For the image reconstruction algorithm and for any
software designed to display or analyze the results, the different
images are distinguished by their names (given by their
\KEYWORD{HDUNAME} keyword). The complete name \KEYWORD{HDUNAME} of the final image is set by the algorithm and must begin with `IMAGE-OI` followed by the value of \KEYWORD{HDUPREFX} given in  \STRING{IMAGE-OI~INPUT~PARAM} if any. This \KEYWORD{HDUPREFX} keyword can be used by the caller to prevent identical HDUNAME conflicts.

In order to continue the iterations of a previous reconstruction run, the
image reconstruction algorithm may be started with the final image instead
of the initial one.  To that end, there must be some means to specify the
starting image for the reconstruction.  This is the purpose of the
\KEYWORD{INIT\_IMG} keyword (see Table~\ref{tab:input-params}) in the input
parameters HDU which indicates the \KEYWORD{HDUNAME} of the initial
image. This \KEYWORD{HDUNAME} must be unique within the file.

\subsection{Model of the data}

As each imaging
algorithm may implement its own method for estimating the complex visibilities
given the image of the object, it is necessary that the model of every fitted
data point be computed by the algorithm itself rather than by another tool.

The OIFITS format \cite{Pauls_et_al-2005-oifits} specifies that
optical interferometric data is stored in binary tables as columns
with specific names.  As there is no restriction that the tables only
contain the columns specified by the standard, we propose to store the
model of the data in the same tables by adding new columns. The
additional columns have the prefix \STRING{NS\_MODEL\_} to distinguish them from
the columns defined by the OIFITS standard. The names of the new
columns are listed in Table~\ref{tab:model-columns}.  In this way it
is very easy to compare the actual data and the corresponding model
values as computed from the reconstructed image and the instrument
model assumed by the reconstruction algorithm.  Another advantage of
this convention is that the same format can be exploited to store the
values given by model fitting software.

\renewcommand{\ROW}[2]{\texttt{#1} & \texttt{D(\textsl{NWAVE})} & #2 \\}
\renewcommand{\ROWTITLE}[1]{\multicolumn{3}{c}{\textbf{#1}}\\}

\begin{table}
\caption{Colums inserted in OIFITS binary tables to store the values given by
the model.  \texttt{\textsl{NWAVE}} is the number of wavelengths.
\label{tab:model-columns}}
\begin{tabular}{lcl}
\hline
\hline
\ROWTITLE{New columns in \texttt{OI\_VIS} tables}
Label & Format & Description \\
\hline
\ROW{NS\_MODEL\_VISAMP}{Model of the visibility amplitude}
\ROW{NS\_MODEL\_VISAMPERR}{Model of the error in visibility amplitude {(optional)}}
\ROW{NS\_MODEL\_VISPHI}{Model of the visibility phase in degrees}
\ROW{NS\_MODEL\_VISPHIERR}{Model of the error in visibility phase in degrees {(optional)}}
\hline
\hline
\ROWTITLE{New columns in \texttt{OI\_VIS2} tables}
Label & Format & Description \\
\hline
\ROW{NS\_MODEL\_VIS2}{Model of the squared visibility}
\ROW{NS\_MODEL\_VIS2ERR}{Model of the error in squared visibility {(optional)}}
\hline
\hline
\ROWTITLE{New columns in \texttt{OI\_T3} tables}
Label & Format & Description \\
\hline
\ROW{NS\_MODEL\_T3AMP}{Model of the triple-product amplitude}
\ROW{NS\_MODEL\_T3AMPERR}{Model of the error in triple-product amplitude {(optional)}}
\ROW{NS\_MODEL\_T3PHI}{Model of the triple-product phase in degrees}
\ROW{NS\_MODEL\_T3PHIERR}{Model of the error in triple-product phase in degrees {(optional)}}
\hline
\end{tabular}
\end{table}

\section{OImaging: a single graphical user interface for image reconstruction algorithms}

OImaging is a Java graphical user interface developed at JMMC\footnote{Jean Marie Mariotti Center \url{http://www.jmmc.fr/}}. With it, image reconstruction is carried out in three main steps:
\begin{itemize}
  \item after loading an OI-FITS files, it creates and fills the \STRING{IMAGE-OI~INPUT~PARAM} and \STRING{IMAGE-OI~INITIAL~IMAGE} HDUs with the values entered by the user in the GUI input panel;
  \item then the IMAGE-OI OI-FITS file is sent to a server (either locally or remotely at JMMC) that launches the chosen algorithm and performs the reconstruction;
  \item finally the reconstructed image is sent back to OImaging and is shown in the result panel.
\end{itemize}

\subsection{Input panel}

\begin{figure}
  \centering
  \includegraphics[width=\textwidth]{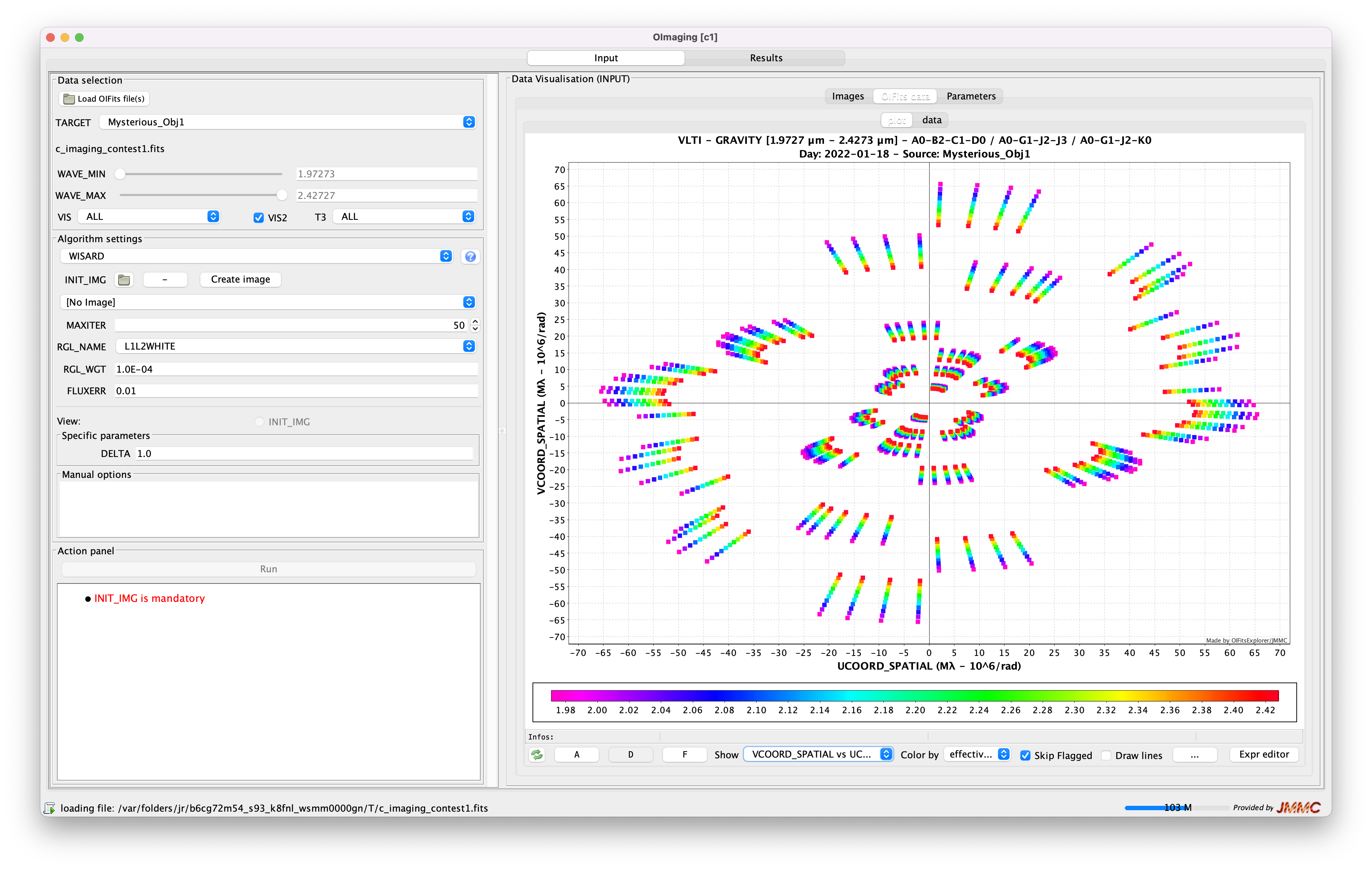}
  \caption{OImaging Input Panel}
\end{figure}

When started OImaging opens on the \texttt{input panel} that is used to load the data, select the algorithm and set the parameters. 

\subsubsection{OIFITS file loading}

The data can be loaded either through the file loader or through the SAMP protocol \cite{taylor2015samp}. This protocol is for sending data directly from one software program to another without saving it to a file. As an example, OIFITS files can be sent from the OIdb database\footnote{\url{http://oidb.jmmc.fr/index.html}} \cite{OIdb} without explicitly saving them locally. If several files are selected they are just merged all together without any kind of filtering. 
When loading an OIFITS file, OImaging will check if it contains an \STRING{IMAGE-OI~INPUT~PARAM} HDU. If so, it will fill the fields of the input panel with the stored values. Otherwise, it will create it with default values. 

\subsubsection{Data visualization}
Loaded data can be explored in the \texttt{visualization sub-panel}. In this sub-panel, the user can plot the data. He can also explore all the values stored in the OIFITS files: all the tables containing the data and in particular the values of the image reconstruction parameters stored in the \STRING{IMAGE-OI~INPUT~PARAM}.
This visualization sub-panel can also be used to display and modify all the images stored in the IMAGE-OI OIFITS file, in particular the initial image and the prior image. 

\subsubsection{Data selection}
As explained above, any sophisticated selection, merging or editing of the data should be done by a separate tool such as JMMC's OIFITSExplorer. However, basic selection tools are available in the input panel, namely the target and the wavelength range. In addition, the user can select the types of data used in the reconstruction from the amplitudes and the phases of the visibilities (VISAMP and VISPHI), the squared visibilities (VIS2) and the amplitudes and the phases of the bispectra (T3AMP and T3PHI).

\subsubsection{Algorithm selection}
The first step for performing image reconstruction is to select the algorithm. At the time of writing, four algorithms are available: WISARD\cite{meimon2005wisard}, BSMEM\cite{baron2008BSMEM}, MiRA\cite{MiRA} and SPARCO\cite{kluska2014sparco}. For each algorithm a small help window can show a brief description, the reference paper and a description of all supported parameters. 

\subsubsection{Parameter settings}
Once the algorithm has been selected, OImaging automatically updates the input fields used to set the parameters. There are three groups of parameters: 
\begin{itemize}
    \item first the parameters shared by all algorithms, namely the number of iterations \KEYWORD{MAXITER}, the regularization name \KEYWORD{RGL\_NAME}, the regularization weight \KEYWORD{RGL\_WGT} and the error on the total flux \KEYWORD{FLUXERR};
    \item the second group contains parameters specific to the algorithm and/or the regularization function;
    \item in addition, there is a text field for \texttt{Manual options} that will be appended to the command line used to invoke the reconstruction software. This is a convenient way to access unsupported features that may change the algorithm behavior or the log contents. It is mainly intended for debugging.
\end{itemize}

\subsubsection{Initial and prior images}

All algorithms need an initial guess of the image. As explained in Sec. \ref{sec:initial}, the pixel size and the dimensions of the
initial image determine those of the reconstructed image. This initial image can be specified by several means:
\begin{itemize}
  \item loading a FITS file locally,
  \item sending the image from another software like SAO-DS9\cite{joye2003sao} or LITpro\cite{tallon2008litpro}. This is particularly convenient as an initial estimate of the target morphology can be done with the model fitting software LITpro. LITpro can then generate the modeled image and send it to OImaging as an initial guess;
  \item OImaging has a small widget accessible by the \texttt{create image} button to create an image of a Gaussian. Depending of the Gaussian width, it can approximate either a Dirac or a constant background. In this widget, OImaging suggests default sizes of both the field of view and the pixel according to the size of the telescope and the maximum baseline;
  \item after some reconstruction iterations, previous results can also be used as initial image; also
  \item any loaded image can be edited to change is pixel size or its field of view by the mean of the \texttt{modify image} button at the top right of the image.
\end{itemize}

Some regularization functions require a prior image, which can be viewed using the \texttt{view} radio-button next to the parameter field. By default, if no prior image is set, the initial image is also used as the prior image.

\subsubsection{Running the algorithm}

When all parameters are correctly set, the algorithm can be run by clicking on the \texttt{Run} button. This button stays grayed out when there is an issue with the parameters. In that case, the user must read the message in the box below it. A second click on the \texttt{Run} button while the reconstruction is in progress cancels the computation.

\subsection{Result panel}

\begin{figure}
  \centering
  \includegraphics[width=\textwidth]{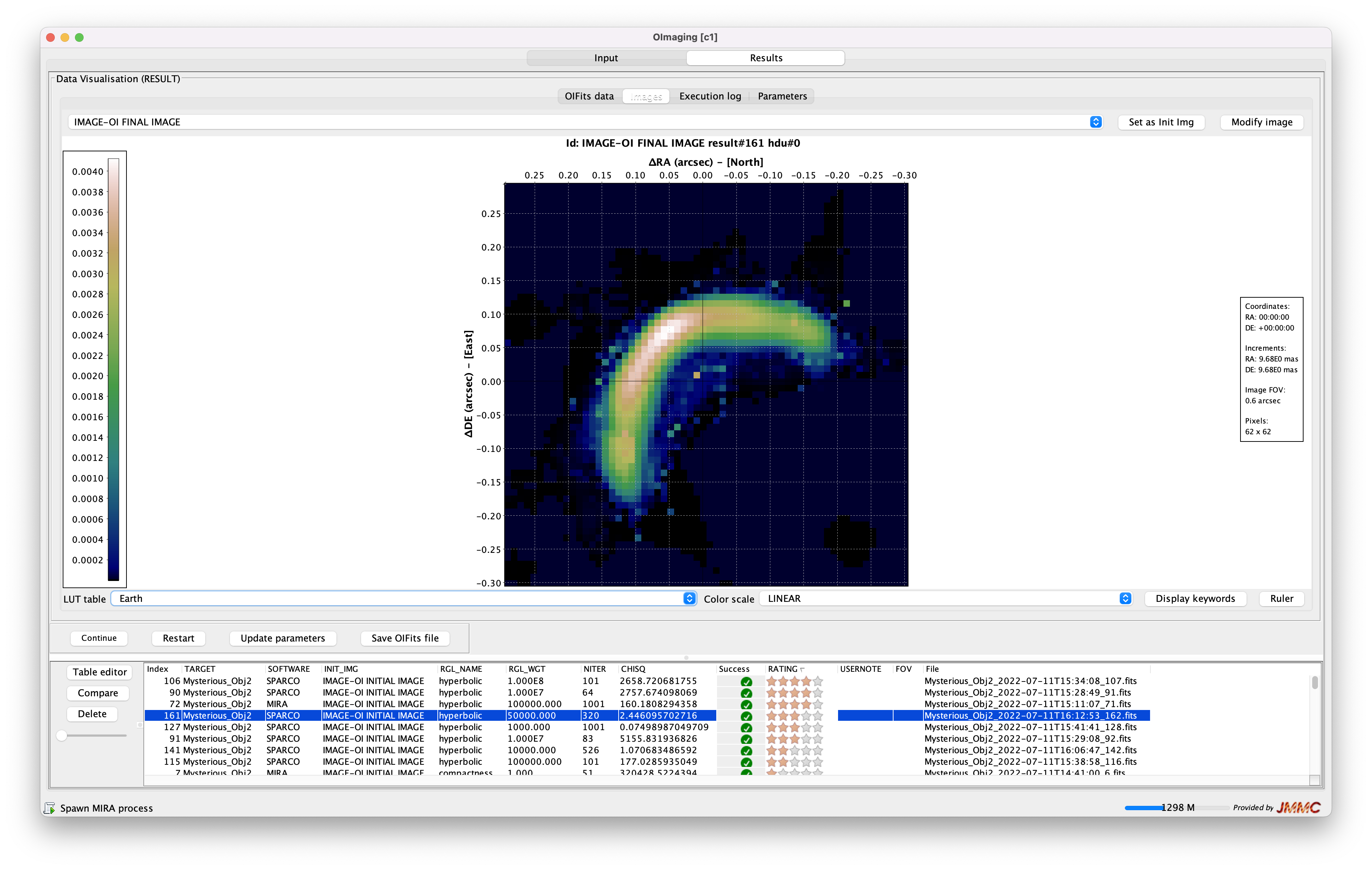}
  \caption{OImaging Result Panel}
\end{figure}

The reconstructed images are shown in the result panel. This panel is composed of two sub-panels: the visualization sub-panel and the results table. 
% References

\subsubsection{Results table}

All results are stored in the results table at the bottom of the result panel. This table gives an overview of all the parameters stored in the result OI-IMAGE OIFITS files: either the input parameters such as the algorithm or the regularization weight or the output parameters computed by the algorithm such as the $\chi^2$ or the value of the regularization function. The table contains  additional fields such as ``Success'' that show if the software stopped with an error. It also contains \KEYWORD{RATING} and \KEYWORD{USERNOTE} fields to allow the user to add additional information concerning a given reconstruction. These fields are all stored in the resulting OIFITS file.

To help the user navigate the results, the table can be sorted and edited. Using the \texttt{Table Editor} button, the table is fully customizable: any keyword present in the OIFITS file can be set as field in the table. Reconstruction results can be also be removed from the table by means of the \texttt{Delete} button.

When several results are selected, the \texttt{compare} button displays the reconstructed images side by side.

\subsubsection{Result visualization}

The selected result in the panel can be explored in a visualization sub-panel similar to the visualization sub-panel of the input panel.  This visualization sub-panel is itself composed of four sub-panels:
\begin{itemize}
  \item \texttt{OIFits data} sub-panel shows all the data stored in the OI-FITS tables either as text or as graphs. As the algorithms store the model of the data as additional columns with the prefix \STRING{NS\_MODEL\_}, the modeled data and the residuals can also be plotted in this panel.
  \item \texttt{Images} sub-panel can display all the images contained in the OI-FITS file, by default the result image but also the initial image or the prior image depending on the HDU name selected in the menu at the top of the panel. This Image sub-panel also has two buttons at the top: a \texttt{Set as init img} button that put the displayed image in the initial image list in the input panel and a \texttt{modify image} button to change the sizes of the pixels and the field of view. In that case, the modified image does not overwrite the current image but is rather put the initial image list in the input panel. This sub-panel contains also a \texttt{ruler} button to help the user to measure distance and angle on the image.
  \item \texttt{Execution log} sub-panel shows the output log of the algorithm.
  \item \texttt{Parameters} sub-panel shows the parameters set in the HDUs \STRING{IMAGE-OI~INPUT~PARAM} and \newline \STRING{IMAGE-OI~OUTPUT~PARAM}.
\end{itemize}

\subsubsection{Interacting with the results}
In addition to the visualization sub-panel and the results table, the result panel shows four buttons for interacting with the selected results:
\begin{itemize}
  \item \texttt{Continue} uses the reconstructed image as the initial image to perform \KEYWORD{MAXITER} more iterations without changing any other parameters. It is a convenient way to perform a reconstruction for a large number of iterations checking the image every \KEYWORD{MAXITER} iterations.
  \item \texttt{Restart} updates all the parameters in the input panel with the values in  \STRING{IMAGE-OI~OUTPUT~PARAM} and with the same initial image.
  \item \texttt{Update parameters} is equivalent to \texttt{Restart} but using the result image as the initial image in the input panel.
  \item \texttt{Save OIFits files} saves the selected result. 
\end{itemize}

\section{Conclusion}
We have presented in this paper a new   OI-FITS extension: \textbf{IMAGE-OI}. This new standard has two mains purpose: (i) to standardize  communicating with image reconstruction software backends and (ii) to store in a single file along with the reconstructed image, the data and all the parameters that where used in the reconstruction process enabling its reproducibility.  

In a second part, we presented  presents \textbf{OImaging}, a graphical user interface that uses this standard to provide a common frontend for several image reconstruction algorithms. This open source software is freely available on JMMC website\footnote{\url{https://www.jmmc.fr/english/tools/data-analysis/oimaging/}}. Its development is still active and several feature may be added in the future (such that the automatic fit of a simple model to generate the initial image, tools to perform batch processing with different parameters). Support for this software is provided via the network of VLTI Expertise Centers and help can be requested at \url{https://apps.jmmc.fr/feedback/}.

\bibliographystyle{spiebib.bst} % makes bibtex use spiebib.bst

\bibliography{Soulez_et_al-2022-OImaging} % bibliography data in report.bib

\end{document}